\documentstyle[a4,11pt,epsf]{article}
\begin{document}
\newcommand{\be}{\begin{equation}}
\newcommand{\ee}{\end{equation}}
\newcommand{\bea}{\begin{eqnarray}}
\newcommand{\eea}{\end{eqnarray}}
\newcommand{\nn}{\nonumber}
\newcommand{\muh}{\hat\mu}
\newcommand{\dlr}{\stackrel{\leftrightarrow}{D} _\mu}
\newcommand{\vnew}{$V^{\rm{NEW}}$}
\newcommand{\vecp}{$\vec p$}
\newcommand{\dof}{{\rm d.o.f.}}
\newcommand{\vvp}{v_B\cdot v_D}
\newcommand{\dl}{\stackrel{\leftarrow}{D}}
\newcommand{\dr}{\stackrel{\rightarrow}{D}}
\newcommand{\mev}{{\rm MeV}}
\newcommand{\gev}{{\rm GeV}}
\newcommand{\calp}{{\cal P}}
\pagestyle{plain}
\def\equiinf{\lower 0.12cm\hbox{$ \widetilde{_{_{t \rightarrow
\infty}}}$ }}
\def\equizer{\lower 0.12cm\hbox{$ \widetilde{_{_{a \rightarrow 0}}}$ }}
\begin{flushright}
 LPTHE Orsay-94/79\\
 hep-ph/9409338
\end{flushright}
\begin{center}
{\Large{Lattice results for Heavy-Light matrix elements}}\\

\vspace*{1cm}
{As. Abada$^{\dag}$ }\\

{$^{\dag}$Laboratoire de Physique Th\'eorique et Hautes Energies} \\
{Universit\'e de Paris XI}
{91405 Orsay FRANCE\footnote {Laboratoire
associ\'e au Centre National de la Recherche Scientifique.}}
\vspace*{1cm}
\end{center}
\abstract{\noindent{We review recent lattice computations relevant for $D$ and
$B$ decays. Decay constants $f_{D,D_s}$, $f_{B,B_s}$, $D\rightarrow K (K^*)$,
$B\rightarrow \pi , \rho$ semi-leptonic form factors together with the slope of
the Isgur-Wise function calculations are presented. Some recent results of
$B\rightarrow K^* \gamma$ form factors will be given. $1/M$ corrections to the
asymptotic scaling laws are discussed.}}

\section{ {Introduction}}
$B$ meson physics is now an active field (where several thousand of physicists
are engaged) in which lattice QCD takes advantage from recent improvements
increasing the statistics and improving the control of the systematics through
theoretical progress (improved actions$\cdots$). In addition to
phenomenological predictions, lattice QCD can test the scaling laws predicted
by HQET (Heavy Quark Effective Theory). Up to now, precise predictions can be
made only in the quenched approximation for which the systematic errors cannot
be evaluated; however, results obtained in the unquenched theory\cite{HEMCGC}
suggest that the quenching effect is small when one deals with heavy quarks. In
general, comparison with experiment shows an agreement with lattice data inside
error bars which are still sizable although decreasing. This is a review of
recent calculations done in the quenched approximation using the Wilson and
clover (continuum limit improved) actions on heavy-light meson decays.
\section{{Strategy to study $B$ meson on the lattice}}
Since the inverse lattice spacing $a^{-1}$ ranges from $2$ to $4$ $\gev$, one
cannot study $B$ meson directly on the lattice. Indirect informations can be
obtained through the following:\par
{\bf-} {\noindent One uses a set of relatively heavy mesons with masses up to
$0.7 a^{-1}$, i.e. heavier than the $D$ but lighter than the $B$ meson (``
fictitious $D$ mesons"). We call this mass region the ``moving quark"
region.}\\
{\bf-} {On the other hand, a method proposed by Eichten\cite{eichten} allows
one to put a quark with infinite mass on the lattice and the latter is
considered in this approach as a static source of color. We call this mass
region the ``static quark" one (where heavy flavours are studied at lowest
order in ${1\over M }$ expansion, $M$ being the heavy meson mass).}\par
A physical quantity computed in these two mass regions is interpolated to the
$B$ meson with the help of the scaling laws of the HQET. The value in the
static limit reduces the uncertainties due to the extrapolation. This method
has shown to be very effective (Fig.1) in the estimation of the decay
constants. In semi-leptonic decays, the calculation in the static limit is not
yet available, but one can study the scaling behaviour and try an extrapolation
to the $B$ meson. The predictions concerning the $B$ meson semi-leptonic decays
remain at a semi-quantitative level but it shows that the extrapolation may be
done and improvements in the near future are expected. The next section is an
illustration of the method explained above.
\section{Leptonic decays $D(B)\rightarrow \ell\nu$}
A hadron mass can be obtained from the study of an appropriate Euclidean
correlation function as the coefficient of its exponential time dependence:
\def\={\ =\ }
\def\x{ {\bf x}}
\def\y{ {\bf y}}
\def\0{ {\bf 0}}
\bea
G(t) &=& \int {\rm d}^3 {\bf x}
\langle \bar u(\x ,t)\gamma_0 \gamma_5 c(\x ,t)
 \bar c(\0 ,0) \gamma_0 \gamma_5 u(\0 ,0) \rangle \nn \\
 &{_{\simeq\atop{t \rightarrow \infty}}}& {f_D^2 m_D\over 2 } {\rm e}^{-m_D t}
\label{GdeT}\eea
The determination of the expectation value in eq.\ref{GdeT} is a non
perturbative problem which can be solved numerically.
The second approach (static) is based on the expansion of the heavy
quark ($H$) propagator in inverse powers of the quark mass as proposed by
Eichten\cite{eichten}; the $H$ is static and does not live effectively
on the lattice but the quantity $f_{_H}\sqrt{M_{_H}}$ can be measured and is
predicted to be independent of the heavy mass. The confrontation between the
two methods is presented in Fig.1. The HQET tells us that when $M_H\rightarrow
\infty$, the vector (V) and pseudoscalar (P) decay constants scale with the
mass of the heavy quark, $M_H$, \cite{eichten}-\cite{voloshin}($M=M_P=M_V=M_H$,
$\beta_0=11-{2\over 3}N_f$) as: \be
{M\over f_V}=f_P={C\over \sqrt{M}}\alpha_s(M)^{-2/\beta_0}.\ee
$$
\epsfbox{moriondfig.ps}
$$
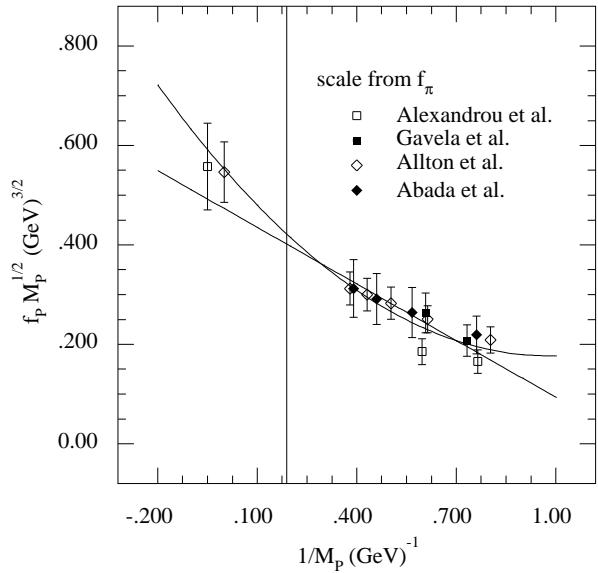
\parbox [t]{\textwidth} {{Figure.1. {\it{Linear and quadratic fit in $1/M$ are
shown,
the vertical line shows the physical $B$ meson.}}}}
\vskip 0.5cm
In Figure.1, we notice the consistency between the moving quark results and the
static ones. It appears that there are large corrections to the asymptotic
scaling behaviour (2). The lattice results compared to the experimental ones
are reported in Table 1(2) concerning the $D$($B$) meson.  \vskip 1cm

\begin{table}
\centering
\begin{tabular}{|c|c|c|c|c|}
\hline
Ref.&$\beta$&$f_D(\mev)$&$f_{D_s}(\mev)$ \\ \hline
ELC\small{(W)}\cite {orsw} &6.4&$210 \pm 15$&$227 \pm 15$\\ \hline
APE \small{(C)}\cite{ape} &6.0&$218\pm 9$ &$240 \pm 9$\\ \hline
BLS\small{(W)}\cite{bernard}&6.3&$208(9)\pm 35\pm 12$&$230(7) \pm 30 \pm 18$
\\ \hline
UKQCD \small{(C)}\cite{ukqcd}&6.2&$185^{+4}_{-3}$ $^{+42}_{-7}$ &$212^{+4}_{-4}
$$^{+46}_{-7}$  \\ \hline
WA75\cite{wa75}&&-&$232 \pm 45 \pm 20 \pm 48$  \\ \hline
CLEO2\cite{cleo2}&&-&$344 \pm 37\pm 52 \pm 42$ \\ \hline
ARGUS\cite{argus}&& &$267 \pm 28$\\ \hline
\end{tabular}
\caption{\it{{W (C) refers to Wilson (clover) action.}}}
\end{table}
\vskip -1cm From Table 1, one can see that the different lattices agree more or
less. Up to $5\%-10\%$ we find: $f_D\sim 210\mev$ and $f_{D_s}\sim 230\mev$.
$f_{D_s}$ which has been predicted by lattice since several years, will provide
an important check since the large experimental errors may be substantially
reduced in the future.
\vskip 0.1cm
\begin{table}
\centering
{
\begin{tabular} {|c|c|c|c|c|}
\hline
Ref.&$\beta$& $f_B$($\mev$)&${f_{B_s}\over f_{B_d}}$\\ \hline
ELC\small{(W)}\cite{orsw} &6.4&$205 \pm 40$&$1.08 \pm 0.06$    \\ \hline
APE \small{(S-C)}\cite{apestat} &6.2&$290\pm 15 \pm 45$&$1.11(3)$ \\ \hline
APE \small{(S)}\cite{apeanc} &6.0&$350\pm 40 \pm 30$&$1.14(4)$ \\ \hline
APE \small{(S-C)}\cite{apeanc} &6.0&$328\pm 36 $&$1.19(5)$ \\ \hline
UKQCD \small{(C)}\cite{ukqcd} &6.2&$160 ^{+6}_{-6}$ $^{+53}_{-19}$&$1.22\pm
0.04$ \\ \hline
UKQCD \small{(S-C)}\cite{ukqcd}&6.2&$253 ^{+16}_{-15}$ $^{+105}_{-14}
$&$1.14^{+4}_{-3}$ \\ \hline
BLS \small{(S)}\cite{bernard}&6.3&$235(20)\pm 21$&$1.11 \pm 0.05 $\\ \hline
Allton\small{(S-W)}\cite{alton} &6.0&$310 \pm 25\pm 50$ &$1.09\pm 0.04$  \\
\hline
HEMCGC\cite{HEMCGC}&5.6&$200\pm 48$&- \\  \hline
\end{tabular}
\caption{\it{S refers to Static limit. The HEMCGC unquenched result agree with
those obtained in the quenched approximation.}}
}
\end{table}
\vskip 0.1cm
In Table 2, the general tendency is $f_B\sim 200\mev$ (up to $20\%$), in
agreement with QCD Sum Rules calculations.\vskip 0.5cm
\noindent {\underline{{The $B-$parameter}}}: the predictions for the $B-\bar B$
mixing depend on the $B-$parameter of the heavy light $\Delta B =2$ four quark
operator. ELC gives\cite{orsw}: ${B_{_{D^0}}} = 1.05 \pm 0.08$, $B_{_{B^0}} =
1.16\pm 0.07$ and
${B_{_{D_s}}\over B_{_{D_d}}} \simeq {B_{_{B_s}}\over B_{_{B_d}}}=1.02 \pm
0.02$; the prediction for the physically relevant combination in $B-\bar B$
mixing and $CP$ violation is: ${f_{B_d}\sqrt{{\rm B}_{B_d}}= 220 \pm 40 \mev}$.
{\section{Semi-leptonic decays $D(B)\rightarrow K,K^* (\pi, \rho) \ell\nu$}}
In this study, we first calculate $D\rightarrow K,K^*$ then extrapolate to
$B\rightarrow \pi, \rho$. The amplitudes are expressed in terms of four form
factors $f^+$, $V$ and $A_{1,2}$ and we need to know them at momentum transfer
$q^2=0$ (where we have the maximum of phase space). The present lattices run in
the close vicinity of $q^2=0$ for the $D$ meson, far from so for the $B$ one.
In the latter case we proceed in two steps:\par
{\bf -} The first is to extrapolate to the $B$ meson mass at fixed momentum
near the no recoil point ($q^2_{\rm {max}}$) and this is doable with the help
of HQET: when $M\rightarrow \infty$ at fixed $\vec
q$ and $||\vec q||\ll M$, the form factors scale as\cite{scala}:
$f^+,V,A_2\sim M^{1/2}$, and $A_1\sim M^{-1/2}$).\par
{\bf -} The 2$^{\rm nd}$ and difficult step is to extrapolate to $q^2=0$;
people often use the Nearest Pole Dominance approximation VMD
($F(q^2)={F(0)\over 1-q^2/M^2_t},M_t$ is the exchanged meson mass in the
t-channel) which has no firm theoretical grounding.\vskip 0.5cm
\noindent {\underline{{$D$ meson study}}}: lattice results compared to model
predictions and experimental data are reported in Table 3; the two first
lines\cite{slape},\cite{slukqcd} have been recently obtained using the clover
action. There is agreement between lattices, models and experiments for $f^+$
and $A_1$. For $V$ the central value is below within errors. Concerning $A_2$,
there used to be a problem but now the central value agrees with the
experimental average although with large errors.\vskip 0.5cm
\noindent {\underline{{$B$ meson study}}}: ELC\cite{semi} and APE\cite{slape}
have studied the HQET scaling laws near $q^2_{max}$ and it seems that there are
large corrections for $V$ and $A_{1,2}$. The $B$ meson form factors predictions
at $q^2=0$ have large errors and rely on the nearest pole dominance
assumption\cite{semi},\cite{slape}. This type of extrapolation has been done
for the first time and now, one knows how to handle ${1\over M}$
corrections.\vskip 0.5cm
\begin{table}
\centering
{{
\begin{tabular}{|c|c|c|c|c|}
\hline
\multicolumn{1}{|c|}{Ref.} &f$^+$(0) &  V(0) & A$_1$(0) \\
\hline \hline
EXP\cite{ave} &$0.77(4)$ & $1.16 \pm .16$ & $0.61(5)$ \\ \hline
 APE\cite{slape} &$0.72(9)$ & $1.00 \pm 0.20$ & $0.64 \pm 0.11$ \\
UKQCD\cite{slukqcd} &$0.67^{+7}_{-8}$ & $0.98^{+10}_{-13} $
&$0.70^{+5}_{-10}$\\
ELC\cite{semi} &$0.60 \pm 0.15 \pm 0.07$&$0.86 \pm 0.24$& $0.64 \pm 0.16$ \\
APE\cite{victor}&$0.63 \pm 0.08$&$0.86 \pm 0.10$&$0.53 \pm 0.03$ \\
BLS\cite{bes}&$0.90\pm 0.08\pm 0.21$& $1.43\pm 0.45\pm 0.49$ &$0.83\pm 0.14\pm
0.28$ \\ \hline
SR\cite{bbd}&$0.60^{+0.15}_{-0.10}$&$1.10 \pm 0.25$&$0.50 \pm 0.15$ \\ \hline
QM.1\cite{wsb}&$0.76$ & $1.23$ &$0.88$ \\
QM.2\cite{wisg} & $0.8$ & $1.1$ & $0.8$ \\ \hline \hline
\multicolumn{1}{|c|}{Ref.} &A$_2$(0) &  V(0)$/$A$_1$(0) & A$_2$(0)$/$A$_1$(0)
\\
\hline \hline
EXP\cite{ave} &
$0.45(9)$ & $1.90 \pm 0.25$ & $0.74 \pm 0.15$ \\ \hline
APE\cite{slape}&$0.46 \pm 0.34$&$1.59 \pm 0.29$&$0.73 \pm 0.45$ \\
UKQCD\cite{slukqcd}&$0.68^{+11}_{-17}$& & \\
ELC\cite{semi}&$0.40 \pm 0.28 \pm 0.04$&$1.3 \pm 0.2$& $0.6 \pm 0.3 $ \\
APE\cite{victor}&$0.19 \pm 0.21$ & $1.6 \pm 0.2$ & $0.4 \pm 0.4$ \\
BLS\cite{bes}&$0.59 \pm 0.14\pm 0.24$ & $1.99 \pm 0.22 \pm 0.33$ & $0.7 \pm
0.16 \pm 0.17$ \\ \hline
SR\cite{bbd} &
$0.60 \pm 0.15$ & $2.2 \pm 0.2$ & $1.2 \pm 0.2$ \\ \hline
QM.1\cite{wsb} &
$1.15$ & $1.4$ & $1.3$ \\
QM.2\cite{wisg} &
$0.8$ & $1.4$ & $1.0$ \\ \hline
\end{tabular}}}
\caption{\it{$D \rightarrow K,K^*$ semi-leptonic form factors. EXP, LAT, QM and
SR refer to experimental average, lattice, quark model and QCD sum rules
calculations respectively.}}
\end{table}

{\underline{{The Isgur-Wise function $\xi(x)$}}}: when the final meson is
heavy, semi-leptonic form factors are expressed in terms of one universal
function, unknown apart from the no recoil point $\xi(1)=1$. In practice we try
to measure its slope around this point $\rho^2=-\xi' (1)$. In the $D$ mass
region, UKQCD found $\rho^2=1.2^{+7}_{-3}$\cite{roukqcd} while CLEO2 gives
$\rho^2=1.01\pm 0.15 \pm 0.09$\cite{roCLEO2}. A preliminary analysis
(UKQCD-Martinelli\cite{unofficial}) gives $\rho^2=1.7\pm 0.2$ for masses $m \ge
m_{D_s}$.

{\section {Radiative Decay ${B}\rightarrow { K}^* \gamma$  }}
The decay rate is expressed in terms of $T_{1,2}(q^2)$ where for a real photon
($q^2=0$), we have the exact condition $T_1(0)=T_2(0)$; the method used to
extract $T_{1,2}$ is the same as the semi-leptonic one:\par
- Use the HQET scaling rules near $q^2_{max}$ (at leading order,
$T_1$$\sim$$\sqrt M$, $T_2$$\sim$${1\over \sqrt M}$) to extrapolate to $M_B$.
With the clover action, APE($\beta =6.0$)\cite{bkstarape} and UKQCD($\beta
=6.2$)\cite{bkstarukqcd} find for $T_2(q^2_{max})$: $0.21(2)$ and
$0.269^{+17}_{-9}$ respectively.\par
- The problem is: how to extrapolate to $q^2=0$? If we apply VMD on the two
factors simultaneously, we find that they scale differently at $q^2=0$
($T_1(0)\sim M^{{-1/2}}$ and $T_2(0)\sim M^{{-3/2}}$) in contradiction with the
fact that they must be equal at $q^2=0$. So for which one is VMD better?

* Applying VMD on $T_2$, APE\cite{bkstarape}, UKQCD\cite{bkstarukqcd} and
Bernard et al.,\cite{bkstarsoni} find for $T_2(0)$: $0.084(7)$,
$0.112^{+7+16}_{-7-16}$ and $0.10\pm 0.01\pm 0.03$ respectively.

* Applying VMD on $T_1$, APE\cite{bkstarape} finds a larger value
$T_1(0)=0.20(7)$. \vskip 0.5cm

There is thus a contradiction between the two approaches; it seems to us  that
the higher value is favoured because indications from lattices (APE and UKQCD)
show that the $q^2$ dependence of $T_2$ is much weaker than would be predicted
by VMD. Using the $1^{\rm st}$ approach, UKQCD finds ${\rm BR}({\rm
B}\rightarrow {\rm K }^*\gamma)=(1.7\pm 0.6({\rm stat})^{+11}_{-9}({\rm
sys})\times 10^{-5}$ while APE, when applying VDM on $T_1$ finds a value closer
(preliminary) to CLEO\cite{bkstarcleo} result: ${\rm BR}=(4.5\pm 1.5\pm
0.9)\times 10^{-5}$.
{\section{Conclusion}}
\noindent There have been good quantitative studies for $f_{D_s}$ ($\sim
230\mev$), $f_{D}$ ($\sim 210\mev$), $f_{B}$ ($\sim 200\mev$ ($20\%$)) and for
$D\rightarrow K^{(*)}\ell\nu$ (agreement up to $10\%$). The study of ${1\over
M}$ corrections (which are found to be large) to HQET are now under control.
Concerning the $B\rightarrow \pi(\rho)\ell\nu$ and $B\rightarrow K^* \gamma$,
since the $q^2$ behaviour is still largely unknown, the predictions at $q^2=0$
are only qualitative.\par
\noindent{\bf{Acknowledgments}

\noindent{\it{This work was supported in part by the CEC Science Project
SC1-CT91-0729 and Human Capital and Mobility Programme, Contract
CHRX-CT93-0132.}}
\vskip 0.2cm
}

\end{document}